\documentclass[11pt,english,a4paper]{article}
\usepackage{times}
\usepackage[T1]{fontenc}
\usepackage[latin1]{inputenc}
\usepackage{geometry}
\usepackage{graphicx}

\makeatletter

\newcommand{\noun}[1]{\textsc{#1}}

\newenvironment{lyxlist}[1]
{\begin{list}{}
{\settowidth{\labelwidth}{#1}
 \setlength{\leftmargin}{\labelwidth}
 \addtolength{\leftmargin}{\labelsep}
 }}
{\end{list}}



\usepackage{babel}
\makeatother
\begin{document}
\begin{flushleft}\textbf{\noun{\LARGE burst statistics as a criterion
for imminent failure}}\par\end{flushleft}{\LARGE \par}

\textbf{\Large \vskip.2in}{\Large \par}

\noindent Srutarshi Pradhan$^{1}$, Alex Hansen$²$ and Per C. Hemmer$³$

\vskip.1in

\noindent \emph{Department of Physics, Norwegian University of Science
and Technology, N--7491 Trondheim, Norway, $¹$pradhan.srutarshi@ntnu.no,
$²$alex.hansen@ntnu.no, $³$per.hemmer@ntnu.no}

\vskip.2in

\begin{lyxlist}{00.00.0000}
\item [{Abstract:}] \noindent {\small The distribution of the magnitudes
of damage avalanches during a failure process typically follows a
power law. When these avalanches are recorded close to the point at
which the system fails catastrophically, we find that the power law
has an exponent which differs from the one characterizing the size
distribution of all avalanches. We demonstrate this analytically for
bundles of many fibers with statistically distributed breakdown thresholds
for the individual fibers. In this case the magnitude distribution
$D(\Delta)$ for the avalanche size $\Delta$ follows a power law
$\Delta^{-\xi}$ with $\xi=3/2$ near complete failure, and $\xi=5/2$
elsewhere. We also study a network of electric fuses, and find numerically
an exponent $2.0$ near breakdown, and $3.0$ elsewhere. We propose
that this crossover in the size distribution may be used as a signal
for imminent system failure. }{\small \par}
\end{lyxlist}
\vskip0.1in

\begin{flushleft}Key words: Failure, Fiber bundle model, Fuse model,
Burst statistics, Crossover\par\end{flushleft}

\section{\noun{introduction}}

Catastrophic failures \cite{Herrmann,Chakrabarti,Sornette,Sahimi,Pratip}
are abundant in nature: earthquakes, landslides, mine-collapses, snow-avalanches
etc. are well-known examples. A sudden catastrophic failure is a curse
to human society due to the devastation it causes in terms of properties
and human lives. Therefore a fundamental challenge is to detect reliable
precursors of such catastrophic events. This is also an important
issue in strength considerations of materials as well as in construction
engineering.

During the failure process of composite materials under external stress,
avalanches of different size are produced where an avalanche consists
of simultaneous rupture of several elements. Such avalanches closely
correspond to the bursts of acoustic emissions \cite{AE-1,AE-2} which
are observed experimentally during the failure process of several
materials. If one counts all avalanches till the complete failure,
their size distribution is typically a power law. However we observed
recently that if one records avalanches not for the entire failure
process, but within a finite interval, a clear crossover behavior
\cite{Pradhan1,Pradhan2} is seen between two power laws, with a large
exponent when the system is far away from the failure point and a
much smaller exponent for avalanches in the vicinity of complete failure.
Therefore such crossover behavior in the burst statistics can be taken
as a criterion for imminent failure. In this report we discuss the
crossover behavior in two different fracture-breakdown models- \emph{fiber
bundle models} {[}10-18] which describe the failure of composite material
under external load and \emph{fuse models} \cite{Herrmann,Hansen2,Zapperi}
which demonstrate the breakdown of electrical networks. An analytic
derivation of the crossover behavior is given for fiber bundle models
and numerical study confirms similar behavior in a fuse model. Thus
we claim that such crossover behavior can be used as a signal of imminent
failure. A recent observation \cite{kawamura06} of the existence
of a crossover behavior in the magnitude distribution of earthquakes
within Japan has strengthened this claim.

\section{\noun{fiber bundle model}}

Our fiber bundle model consists of $N$ elastic and parallel fibers,
clamped at both ends, with statistically distributed thresholds for
breakdown of individual fibers (Figure\ 1). The individual thresholds
$x_{i}$ are assumed to be independent random variables drawn from
the same cumulative distribution function $P(x)$ and a corresponding
density function $p(x)$: \begin{equation}
\mbox{Prob}(x_{i}<x)=P(x)=\int_{0}^{x}p(u)\; du.\label{1}\end{equation}

\begin{center}\includegraphics[width=2.2in,height=1.8in]{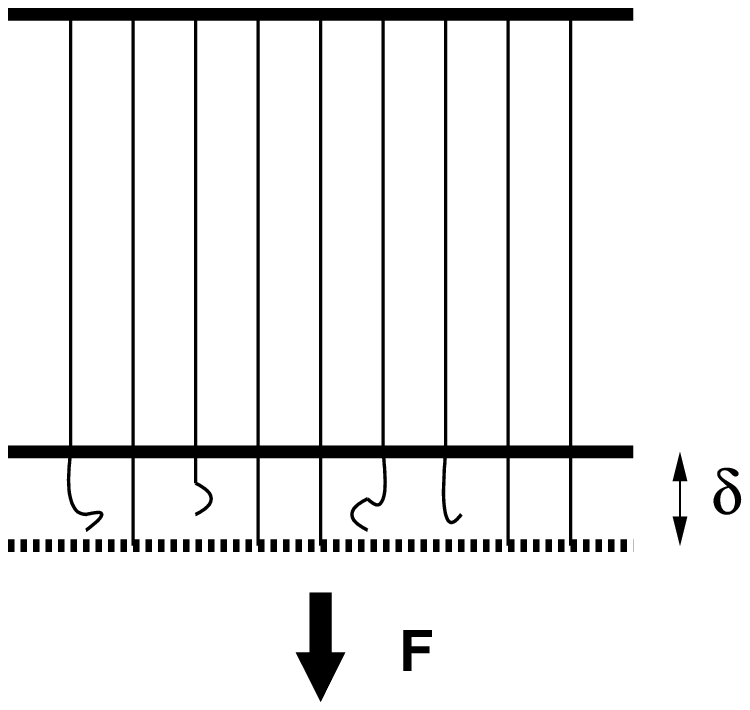}\par\end{center}

\noindent \emph{\small Figure 1}{\small . A fiber bundle of $N$ parallel
fibers clamped at both ends. The externally applied force $F$ corresponds
to a stretching by an amount $\delta$.}\\

Whenever a fiber experiences a force equal to or greater than strength
threshold $x_{i}$, it breaks immediately and does not contribute
to the strength of the bundle thereafter. The maximal load the bundle
can resist before complete breakdown is called the \textit{critical}
load and its value depends upon the probability distribution of the
thresholds. Two popular examples of threshold distributions are the
uniform distribution \begin{equation}
P(x)=\left\{ \begin{array}{cl}
x/x_{r} & \mbox{ for }0\leq x\leq x_{r}\\
1 & \mbox{ for }x>x_{r},\end{array}\right.\label{uniform}\end{equation}
 and the Weibull distribution \begin{equation}
P(x)=1-\exp(-(x/x_{r})^{\kappa}).\label{Weibull}\end{equation}
 Here $x_{r}$ is a reference threshold, and the dimensionless number
$\kappa$ is the Weibull index (Figure\ 2.)

\begin{center}\includegraphics[width=2in,height=1.8in]{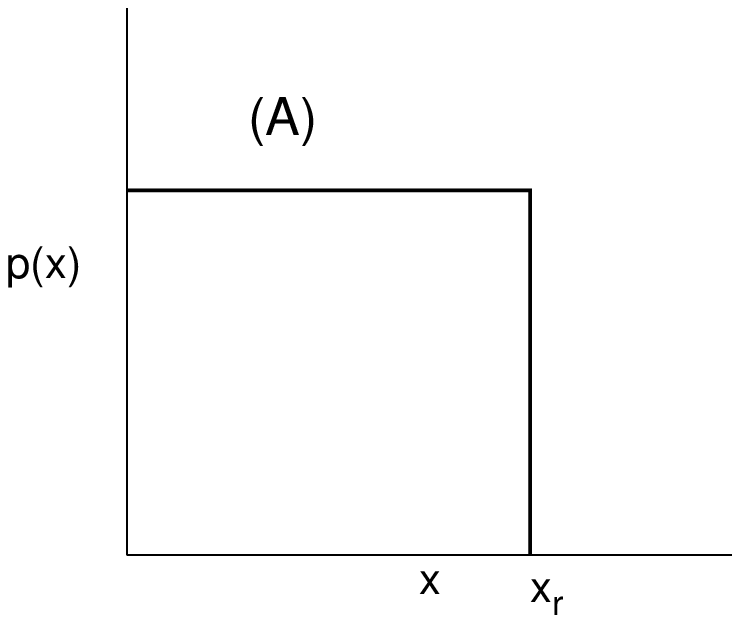}\hskip.2in\includegraphics[width=2in,height=1.8in]{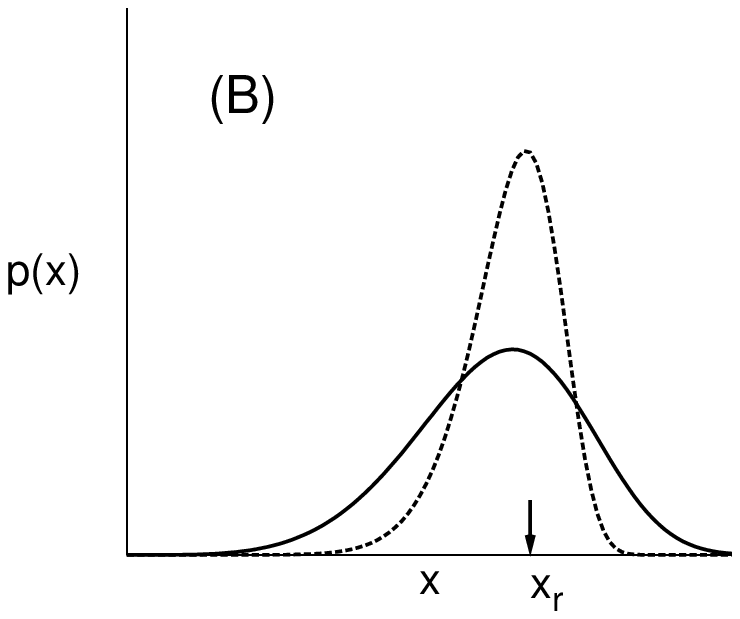}\par\end{center}

\vskip.2in

\noindent \emph{\small Figure 2}{\small . The uniform distribution
(A) and the Weibull distribution (B) with $\kappa=5$ (solid line)
and $\kappa=10$ (dotted line) .}\\

Fiber bundle models differ in the mechanism for how the extra stress
caused by a fiber failure is redistributed among the unbroken fibers.
The simplest models are the equal-load-sharing models, in which the
load previously carried by a failed fiber is shared equally by all
the remaining intact fibers in the system. In the present article
we study the statistics of burst avalanches for equal-load-sharing
models.

\subsection{Burst statistics}

The \textit{burst distribution} $D(\Delta)$ is defined as the expected
number of bursts in which $\Delta$ fibers break simultaneously when
the bundle is stretched steadily until complete breakdown. Hemmer
and Hansen performed a detail statistical analysis \cite{Hemmer}
to find the burst distribution for this quasi-static situation. For
a bundle of many fibers they calculated the probability of a burst
of size $\Delta$ starting at fiber $k$ with threshold value $x_{k}$
as

\begin{equation}
{\displaystyle \frac{\Delta^{\Delta-1}}{\Delta!}\left[1-\frac{x_{k}p(x_{k})}{Q(x_{k})}\right]\left[\frac{x_{k}p(x_{k})}{Q(x_{k})}\right]^{\Delta-1}\times\exp\left[-\Delta\frac{x_{k}p(x_{k})}{Q(x_{k})}\right],}\label{C}\end{equation}
 where $Q(x)=1-P(x)$. Since a burst of size $\Delta$ can occur at
any point before complete breakdown, the above expression has to be
integrated over all possible values of $x_{k}$, i.e, from $0$ to
$x_{c}$ where $x_{c}$ is the maximum amount of stretching beyond
which the bundle collapses completely. Therefore the burst distribution
is given by \cite{Hemmer}

\begin{flushleft}\begin{equation}
\frac{D(\Delta)}{N}={\displaystyle \frac{\Delta^{\Delta-1}e^{-\Delta}}{\Delta!}\int_{0}^{x_{c}}p(x)r(x)[1-r(x)]^{\Delta-1}\exp\left[\Delta\, r(x)\right]dx,}\label{D}\end{equation}
 where \begin{equation}
r(x)=1-\frac{x\, p(x)}{Q(x)}=\frac{1}{Q(x)}\;\frac{d}{dx}\left[x\, Q(x)\right].\label{eq:8}\end{equation}
 \par\end{flushleft}

The integration yields the asymptotic behavior \begin{equation}
D(\Delta)/N\propto\Delta^{-\frac{5}{2}},\label{asymp}\end{equation}
 which is universal under mild restrictions on the threshold distributions
\cite{Kloster}. As a check, we have done simulation experiments for
different threshold distributions. Figure 3 show results for the uniform
threshold distribution (\ref{uniform}) and the Weibull distribution
(\ref{Weibull}) with index $\kappa=5$.

\begin{center}\includegraphics[width=2.2in,height=1.8in]{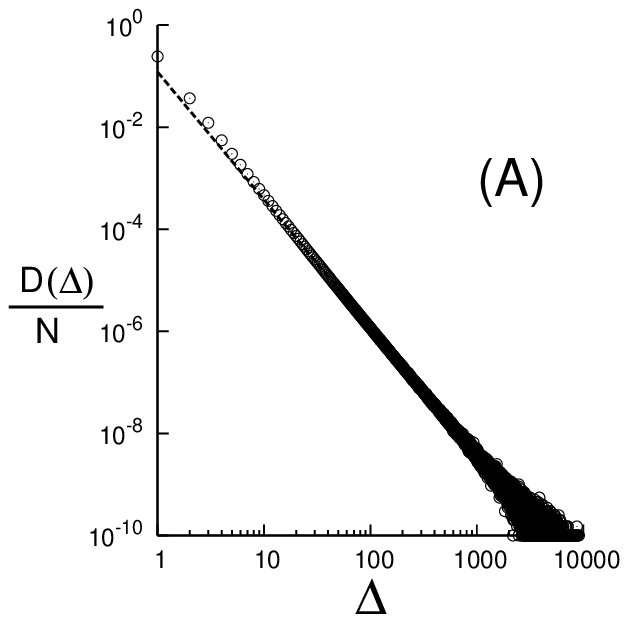}\hskip.2in\includegraphics[width=2.2in,height=1.8in]{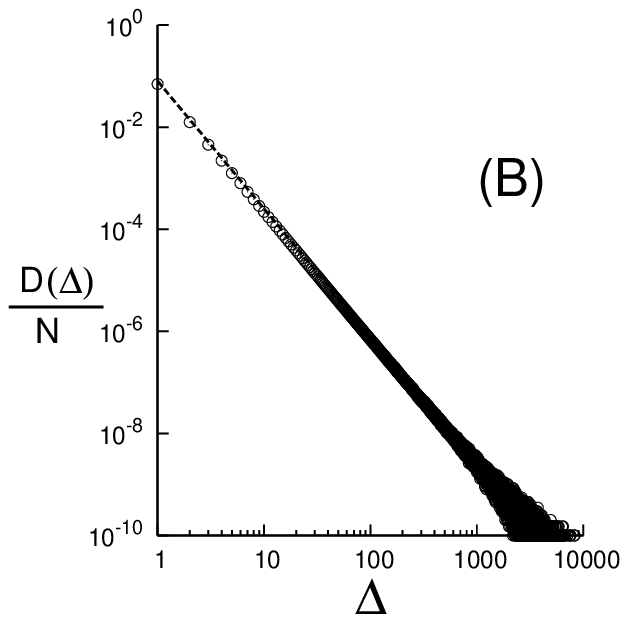}\par\end{center}

\noindent \emph{\small Figure 3}{\small . The burst distribution $D(\Delta)/N$
for the uniform distribution (A) and the Weibull distribution with
index 5 (B). The dotted lines represent the power law with exponent
$-5/2$. Both figures are based on $20000$ samples of bundles each
with $N=10^{6}$ fibers.}\\

\subsection{Crossover behavior near failure point}

\begin{center}\includegraphics[width=2.2in,height=1.8in]{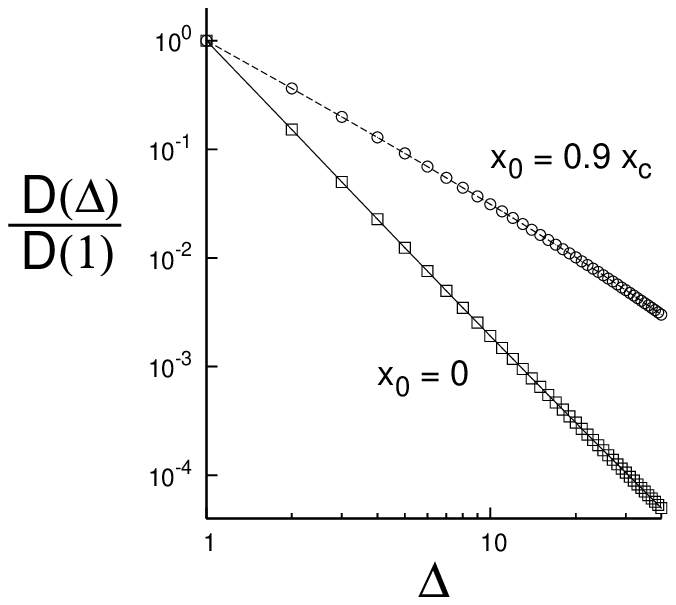}\par\end{center}

\noindent \emph{\small Figure 4}{\small . The distribution of bursts
for thresholds uniformly distributed in an interval $(x_{0},x_{c})$,
with $x_{0}=0$ and with $x_{0}=0.9x_{c}$. The figure is based on
50 000 samples, each with $N=10^{6}$ fibers.}{\small \par}

\vskip.1in

When all the bursts are recorded for the entire failure process, we
have seen that the burst distribution $D(\Delta)$ follows the asymptotic
power law $D\propto\Delta^{-5/2}$. If we just sample bursts that
occur near criticality, a different behavior is seen. As an illustration
we consider the uniform threshold distribution, and compare the complete
burst distribution with what one gets when one samples merely burst
from breaking fibers in the threshold interval $(0.9x_{c},x_{c})$.
Figure\ 4 shows clearly that in the latter case a different power
law is seen.

From Eq. \ref{eq:8} we see that $r(x)$ vanishes at the point $x_{c}$.
If we have a situation in which the weakest fiber has its threshold
$x_{0}$ just a little below the critical value $x_{c}$, the contribution
to the integral in the expression (\ref{D}) for the burst distribution
will come from a small neighborhood of $x_{c}$. Since $r(x)$ vanishes
at $x_{c}${\small ,} it is small here, and we may in this narrow
interval approximate the $\Delta$-dependent factors in (\ref{D})
as follows

\begin{eqnarray}
(1-r)^{\Delta}\, e^{\Delta\, r} & = & \exp\left[\Delta(\ln(1-r)+r)\right]\nonumber \\
 & = & \exp[-\Delta(r^{2}/2+{\mathcal{O}}(r^{3}))]\approx\exp\left[-\Delta r(x)^{2}/2\right]\label{D3}\end{eqnarray}

\noindent We also have \begin{equation}
r(x)\approx r'(x_{c})(x-x_{c}).\end{equation}

\noindent  Inserting everything into Eq.\ (\ref{D}), we obtain to
dominating order \begin{eqnarray}
\frac{D(\Delta)}{N} & = & \frac{\Delta^{\Delta-1}\, e^{-\Delta}}{\Delta!}\int_{x_{0}}^{x_{c}}p(x_{c})\; r'(x_{c})(x-x_{c})e^{-\Delta\, r'(x_{c})^{2}(x-x_{c})^{2}/2}\; dx\nonumber \\
 & = & \frac{\Delta^{\Delta-2}\, e^{-\Delta}p(x_{c})}{\left|r'(x_{c})\right|\Delta!}\left[e^{-\Delta\, r'(x_{c})^{2}(x-x_{c})^{2}/2}\right]_{x_{0}}^{x_{c}}\nonumber \\
 & = & \frac{\Delta^{\Delta-2}\, e^{-\Delta}}{\Delta!}\frac{p(x_{c})}{\left|r'(x_{c})\right|}\left[1-e^{-\Delta/\Delta_{c}}\right],\label{D3}\end{eqnarray}
 with \begin{equation}
\Delta_{c}=\frac{2}{r'(x_{c})^{2}(x_{c}-x_{0})^{2}}.\label{Dc}\end{equation}

By use of the Stirling approximation $\Delta!\simeq\Delta^{\Delta}e^{-\Delta}\sqrt{2\pi\Delta}$,
the burst distribution (\ref{D3}) may be written as\begin{equation}
\frac{D(\Delta)}{N}=C\Delta^{-5/2}\left(1-e^{-\Delta/\Delta_{c}}\right),\label{D2}\end{equation}
 with a nonzero constant \begin{equation}
C=(2\pi)^{-1/2}p(x_{c})/\left|r'(x_{c})\right|.\end{equation}
 We can see from (\ref{D2}) that there is a crossover at a burst
length around $\Delta_{c}$:\begin{equation}
\frac{D(\Delta)}{N}\propto\left\{ \begin{array}{cl}
\Delta^{-3/2} & \mbox{ for }\Delta\ll\Delta_{c}\\
\Delta^{-5/2} & \mbox{ for }\Delta\gg\Delta_{c}\end{array}\right.\end{equation}

We have thus shown the existence of a crossover from the generic asymptotic
behavior $D\propto\Delta^{-5/2}$ to the power law $D\propto\Delta^{-3/2}$
near criticality, i.e., near global breakdown. The crossover is a
universal phenomenon, independent of the threshold distribution $p(x)$. 

\begin{center}\includegraphics[width=2.2in,height=1.8in]{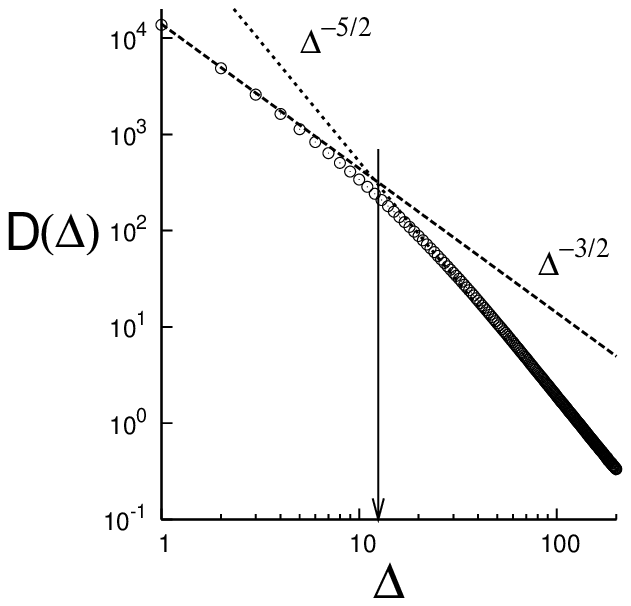}\hskip.2in\includegraphics[width=2.2in,height=1.8in]{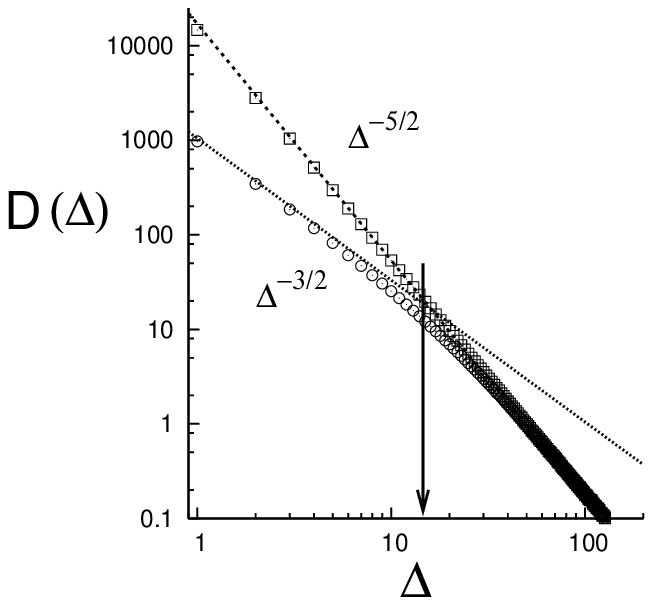}\par\end{center}

\noindent \emph{\small Figure 5}{\small . The distribution of bursts
for the uniform threshold distribution (left) with $x_{0}=0.80x_{\textrm{c}}$
and for a Weibull distribution (right) with $x_{0}=1$ (square) and
$x_{0}=1.7$ (circle). Both the figures are based on $50000$ samples
with $N=10^{6}$ fibers each. The straight lines represent two different
power laws, and the arrows locate the crossover points $\Delta_{c}\simeq12.5$}
{\small and $\Delta_{c}\simeq14.6$,} respectively.

\vskip.1in

For the uniform distribution $\Delta_{c}=(1-x_{0}/x_{c})^{-2}/2$,
so for $x_{0}=0.8\, x_{\textrm{c}}$, we have $\Delta_{c}=12.5$.
For the Weibull distribution $P(x)=1-\exp(-(x-1)^{10})$, where $1\leq x\leq\infty$,
we get $x_{c}=1.72858$ and for $x_{0}=1.7$, the crossover point
will be at $\Delta_{c}\simeq14.6$. Such crossover is clearly observed
(Figure 5) near the expected values $\Delta=\Delta_{c}=12.5$ and
$\Delta=\Delta_{c}=14.6$, respectively, for the above distributions.

\begin{center}\includegraphics[width=2.2in,height=1.8in]{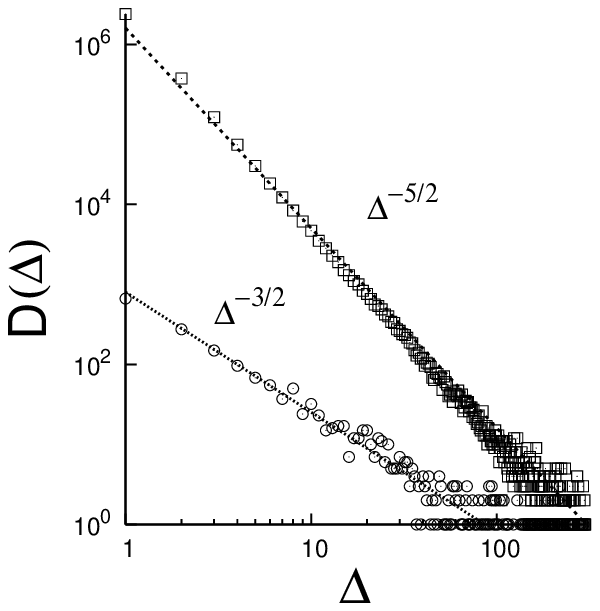}\par\end{center}

\noindent \emph{\small Figure 6}{\small . The distribution of bursts
for the uniform threshold distribution for a single fiber bundle with
$N=10^{7}$ fibers. Results with $x_{0}=0$, i.e., when all avalanches
are recorded, are shown as squares and data for avalanches near the
critical point ($x_{0}=0.9x_{c}$) are shown by circles. }{\small \par}

\vskip.1in

The simulation results we have shown so far are based on \textit{averaging}
over a large number of samples. For applications it is important that
crossover signal can be seen also in a single sample. We show in Figure\ 6
that equally clear crossover behavior is seen in a \textit{single}
fiber bundle when $N$ is large enough. Also, as a practical tool
one must sample finite intervals ($x_{i}$, $x_{f}$) during the fracture
process. The crossover will be observed when the interval is close
to the failure point \cite{Pradhan2}.

\section{\noun{burst statistics in the fuse model}}

\begin{center}\includegraphics[width=2.2in,height=1.8in]{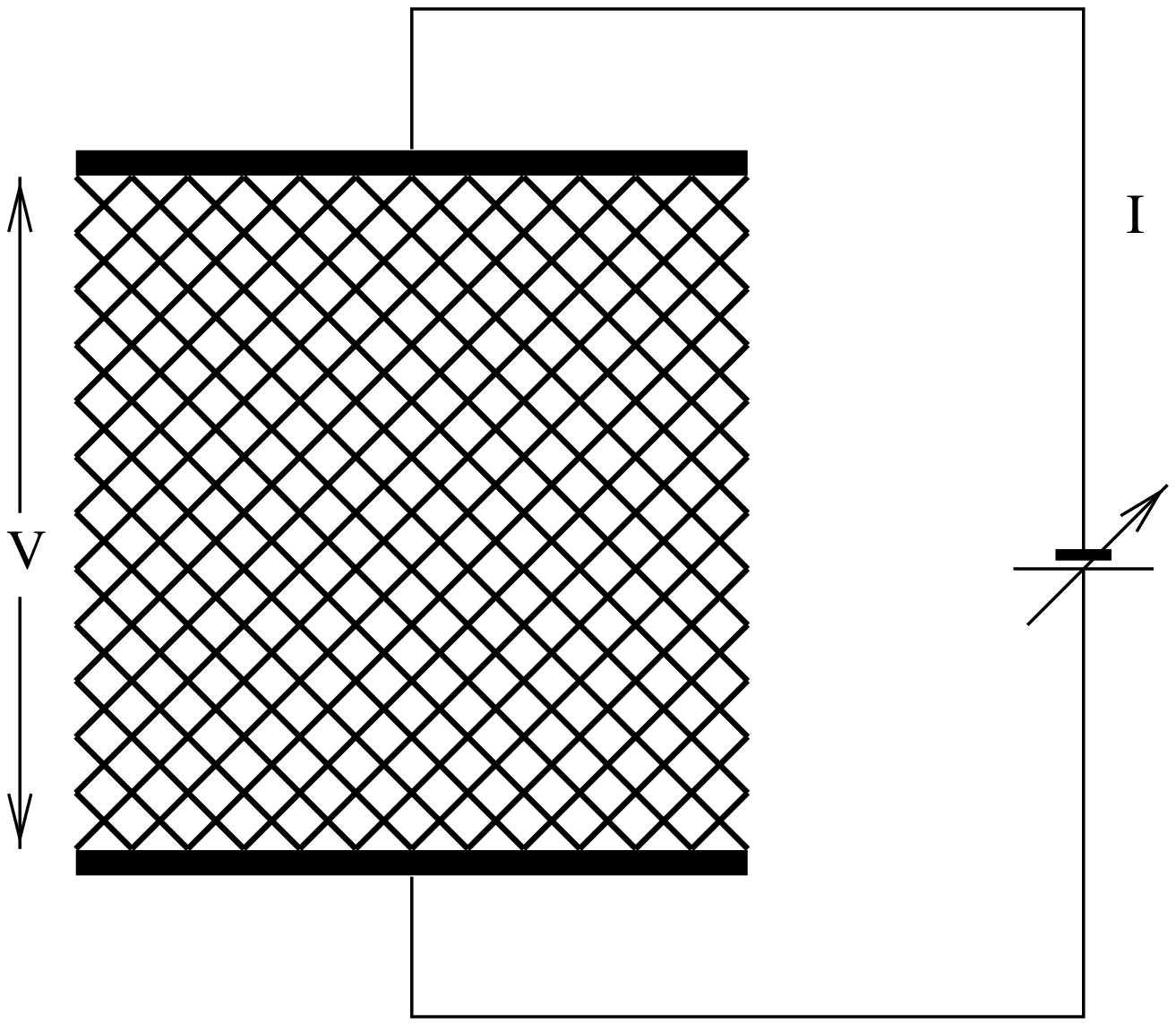}\hskip.6in
\includegraphics[width=2.2in,height=1.8in]{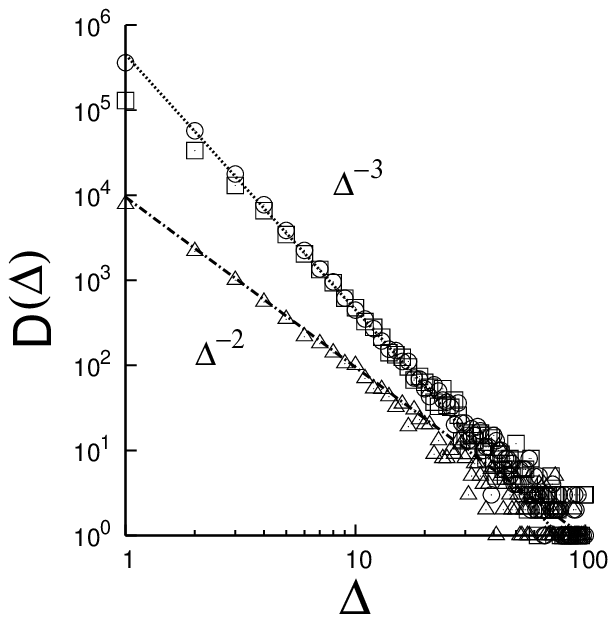}\par\end{center}

\begin{flushleft}\emph{\footnotesize Figure 7}{\footnotesize .} {\footnotesize A
fuse model and the burst distribution: System size is $100\times100$
and averages are taken for $300$ samples. On the average, catastrophic
failure sets in after $2097$ fuses have blown. The circles denote
the burst distribution measured throughout the entire breakdown process.
The squares denote the burst distribution based on bursts appearing
after the first $1000$ fuses have blown. The triangles denote the
burst distribution after $2090$ fuses have blown. The two straight
lines indicate power laws with exponents $\xi=3$ and $\xi=2$, respectively. }\par\end{flushleft}{\footnotesize \par}

\vskip.1in

To test the crossover phenomenon in a more complex situation than
for fiber bundles, we have studied burst distributions in the fuse
model \cite{Herrmann}. It consists of a lattice in which each bond
is a fuse, i.e., an ohmic resistor as long as the electric current
it carries is below a threshold value. If the threshold is exceeded,
the fuse burns out irreversibly. The threshold $t$ of each bond is
drawn from an uncorrelated distribution $p(t)$. The lattice is placed
between electrical bus bars and an increasing current is passed through
it. The lattice is a two-dimensional square one placed at $45{}^{\circ}$
with regards to the bus bars, and the Kirchhoff equations are solved
numerically at each node assuming that all fuses have the same resistance.

When one records all the bursts, the distribution follows a power
law $D(\Delta)\propto\Delta^{-\xi}$ with $\xi=3$, which is consistent
with the value reported in recent studies \cite{Hansen2,Zapperi}.
We show the histogram in Figure 7. With a system size of $100\times100$,
$2097$ fuses blow on the average before catastrophic failure sets
in. When measuring the burst distribution only after the first $2090$
fuses have blown, a different power law is found, this time with $\xi=2$.
After $1000$ blown fuses, on the other hand, $\xi$ remains the same
as for the histogram recording the entire failure process (Figure
7).

In Figure 8, we show the power dissipation $E$ in the network as
a function of the number of blown fuses and as a function of the total
current. The dissipation is given as the product of the voltage drop
across the network $V$ times the total current that flows through
it. The breakdown process starts by following the lower curve, and
follows the upper curve returning to the origin. It is interesting
to note the linearity of the unstable branch of this curve. In Figure
9, we record the avalanche distribution for power dissipation, $D_{d}(\Delta)$.
Recording, as before, the avalanche distribution throughout the entire
process as well as recording only close to the point at which the
system catastrophically fails, result in two power laws, with exponents
{\small $\xi=2.7$} and {\small $\xi=1.9$,} respectively. It is interesting
to note that in this case there is not a difference of unity between
the two exponents. The power dissipation in the fuse model corresponds
to the stored elastic energy in a network of elastic elements. Hence,
the power dissipation avalanche histogram would in the mechanical
system correspond to the released energy. Such a mechanical system
could serve as a simple model for earthquakes.

\vskip.3in

\begin{center}\includegraphics[width=2.2in,height=1.8in]{fig10} \hskip.4in
\includegraphics[width=2.2in,height=1.8in]{fig11}\par\end{center}

\begin{flushleft}\emph{\small Figure 8}{\small . Power dissipation
$E$ as a function of the number of broken bonds (left) and as a function
of the total current $I$ flowing in the fuse model (right). }\par\end{flushleft}{\small \par}

\vskip.1in

The Gutenberg-Richter law \cite{Herrmann,Chakrabarti} relating the
frequency of earthquakes with their magnitude is essentially a measure
of the elastic energy released in the earth's crust, as the magnitude
of an earthquake is the logarithm of the elastic energy released.
Hence, the power dissipation avalanche histogram $D_{d}(\Delta)$
in the fuse model corresponds to the quantity that the Gutenberg-Richter
law addresses in seismology. Furthermore, the power law character
of $D_{d}(\Delta)$ is consistent with the form of the Gutenberg-Richter
law. It is then intriguing that there is a change in exponent {\small $\xi$}
also for this quantity when failure is imminent.

\vskip.3in

\begin{center}\includegraphics[width=2.2in,height=2in]{fig12}\par\end{center}

\noindent \emph{\small Figure 9}{\small . The power dissipation avalanche
histogram} $D_{d}(\Delta)$ {\small for the fuse model. The slopes
of the two straight lines are $-2.7$ and $-1.9$, respectively. The
circles show the histogram of avalanches recorded through the entire
process, whereas the squares show the histogram recorded after $2090$
fuses have blown. }{\small \par}

\section{\noun{concluding remarks}}

Establishing a signature of imminent failure is the principal objective
in our study of different breakdown phenomena. The same goal is of
course central to earthquake prediction scheme. As in different failure
situations, \emph{bursts} can be recorded from outside -without disturbing
the ongoing failure process, burst statistics are much easily available
and also contain reliable information of the failure process. Therefore,
any signature in burst statistics that can warn of imminent system
failure would be very useful in the sense of wide scope of applicability.
The crossover behavior in burst distributions, we found in the fiber
bundle models, is such a signature which signals that catastrophic
failure is imminent. Similar crossover behavior is also seen in the
burst distribution and energy distributions of the fuse model. Most
important is that this crossover signal does not hinge on observing
rare events and is seen also in a single system. Therefore, such signature
has a strong potential to be used as useful detection tool. It should
be mentioned that most recently, Kawamura \cite{kawamura06} has observed
a change in exponent values of the local magnitude distributions of
earthquakes in Japan, before the onset of a mainshock (Figure 10).
This observation has definitely strengthened our claim of using crossover
signals in burst statistics as a criterion for imminent failure.

\begin{center}\includegraphics[width=3in,height=2.7in]{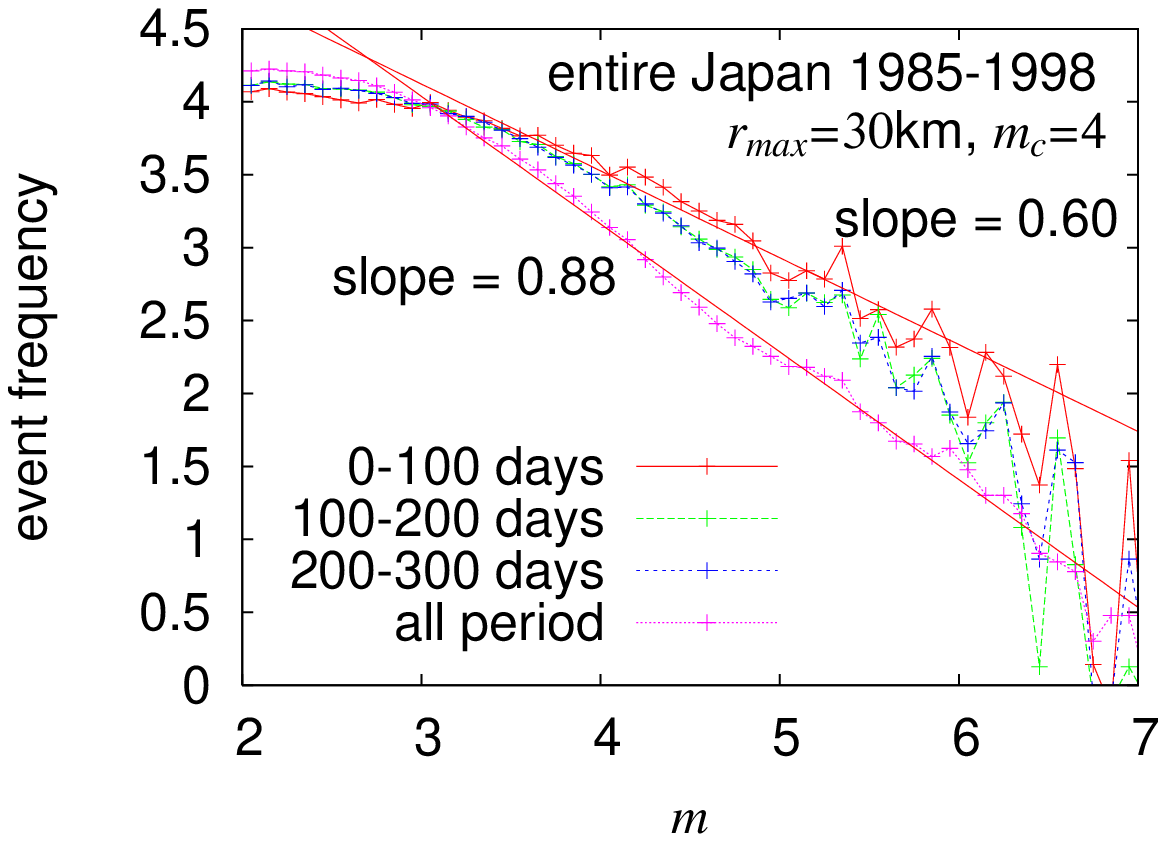}\par\end{center}

\noindent \emph{\small Figure 10}{\small . Crossover signature in
the local magnitude distributions of earthquakes in Japan . The exponent
of the distribution during $100$ days before a mainshock is about
$0.60$, much smaller than the average value $0.88$} {\small \cite{kawamura06}}. {\small }{\small \par}

\vskip.2in

\begin{flushleft}\textbf{\noun{\Large Acknowledgment}}\par\end{flushleft}{\Large \par}

\vskip.2in

S. P. thanks the Research Council of Norway (NFR) for financial support
through Grant No. 166720/V30.

\end{document}